\documentclass[aps,pra,twocolumn,showpacs,groupedaddress,amsmath,amssymb]{revtex4-1}
\usepackage{mathrsfs}

\usepackage{graphicx}
\usepackage{bm}
\usepackage{float}
\begin{document}

\title{Suppression of FM-to-AM conversion in third-harmonic generation \\ by tuning the ratio of modulation depth}

\author{Yisheng Yang$^{1,2}$}
\author{Yizhou Tan$^{1}$}
\author{Bin Feng$^{2}$}
\author{Fuquan Li$^{2}$}
\author{Wei Han$^{2}$}
\author{Jichun Tan$^{1}$}

\affiliation{$^1$College of Science, National University of Defense Technology, Changsha 410073, China\\ $^2$Research Center of Laser Fusion, China Academy of Engineering Physics, Mianyang 621900, China}


\begin{abstract}
Issues of Frequency-to-Amplitude modulation (FM-to-AM) conversion occurred in phase-modulated third-harmonic generation (THG) process are investigated. An expression about group-velocity is theoretically derived to suppress the FM-to-AM conversion, which appears to be dependant on the ratio of modulation depth of fundamental to second-harmonic when given the same modulation frequencies of them. Simulation results indicate that the induced AM in THG process can be suppressed effectively when the expression about group-velocity is satisfied.
\end{abstract}


\maketitle
 Third-harmonic generation (THG) is a powerful technique to produce tunable-wavelength laser pulses, which have varieties of applications in fields such as inertial confinement fusion (ICF), photolithography, and biology \cite{Wegner,Photolith}. To meet the requirement of suppression of stimulated Brillouin scattering or beam smoothing, it is desirable to efficiently frequency-triple laser pulse that has a broad spectrum imposed by phase modulation \cite{Eimerl}. Ideally, this phase modulation does not induce any variations in pulse intensity. However, as result of frequency-dependent effects like group-velocity dispersion, frequency modulation (FM) of input fields will be converted into amplitude modulation (AM) of output fields, which is named as FM-to-AM conversion \cite{Hocquet}. Generally, the induced AM can lead to some higher-order nonlinear effects or may cause damages to optical elements due to instantaneous ultrahigh intensity in process, and thus needs to be prevented \cite{Marozas,Skupsky}. Recently, the suppression of this FM-to-AM conversion has been attracting increasing interests, and many approaches, such as angular spectral dispersion, dual-tripler scheme, and precompensation with gratings or crystals, have been proposed and demonstrated \cite{Hocquetsecond}-\cite{Wangwei}.

 In a previous paper, we reported that the induced AM in THG process could be suppressed at the retracing point of a crystal \cite{Yisheng}. Here, the issue of FM-to-AM conversion in THG process is investigated from the viewpoint of modulation properties of laser pulses. An expression about group-velocity, which reveals the intrinsic group-velocity-matched relationship in phase-modulated THG process, is given to guide the suppression of FM-to-AM conversion.

 Considering the effect of group-velocity mismatch, nonlinear coupling equations describing the THG process under the plane wave approximation can be written as \cite{Armstrong}:
 \begin{equation}
     \frac{\partial{A_{1}(z,t)}}{\partial{z}}+\frac{1}{v_{g1}}\frac{\partial{A_{1}(z,t)}}{\partial{t}}=\frac{i\omega_{1}d_{e}}{n_{1}c}A_{3}A_{2}^{*}e^{i\Delta k_{0}z},
 \end{equation}

 \begin{equation}
     \frac{\partial{A_{2}(z,t)}}{\partial{z}}+\frac{1}{v_{g2}}\frac{\partial{A_{2}(z,t)}}{\partial{t}}=\frac{i\omega_{2}d_{e}}{n_{2}c}A_{3}A_{1}^{*}e^{i\Delta k_{0}z},
 \end{equation}

 \begin{equation}
     \frac{\partial{A_{3}(z,t)}}{\partial{z}}+\frac{1}{v_{g3}}\frac{\partial{A_{3}(z,t)}}{\partial{t}}=\frac{i\omega_{3}d_{e}}{n_{3}c}A_{1}A_{2}e^{-i\Delta k_{0}z},
 \end{equation}
 where subscripts 1, 2, and 3 refer to fundamental (FH), second-harmonic (SH), and third-harmonic (TH) pulse, respectively.
 Parameters $v_{g}$, $\Delta{k_{0}}$, and $d_{e}$ represent group-velocity, original phase-mismatch and nonlinear coefficient, respectively.

 For FH and SH Gaussian pulses with sinusoidal phase modulation, amplitudes can be written as the form of $\exp(-t^{2}/2T^{2}_{i})\exp\big[\sigma_{i}\sin(2\pi\Omega_{i}t)\big]$, where $\sigma$ and $\Omega$ are modulation depth and modulation frequency, respectively. By transforming the coordinate $(z,t)$ to local coordinates $(z,T=t-z/v_{g1})$ and $(z,T'=t-z/v_{g2})$, amplitude of the output TH field in frequency domain can be obtained under the pump undepletion approximation:
 \begin{equation}\label{eq:A3amplitude}
     \tilde{A_{3}}(z,\omega)=\frac{i\omega_{3}d_{e}}{n_{3}c}\int^{z}_{0}\Re\exp(-i\Delta k_{0}\xi)d\xi,
 \end{equation}
 where
 \begin{eqnarray}\label{eq:daR}
     \Re&&=\int^{\infty}_{-\infty}\exp\big[-a(t+\xi\nu_{1})^{2}-b(t+\xi\nu_{2})^{2}\big]\\\nonumber
     &&\cdot\exp\Big[i\sigma_{1}\sin\big(2\pi\Omega_{1}(t+\xi\nu_{1})\big)+i\sigma_{2}\sin\big(2\pi\Omega_{2}(t+\xi\nu_{2})\big)\Big]\\\nonumber
     &&\cdot\exp(i\omega t)d t\nonumber.
 \end{eqnarray}
 In Eq.(\ref{eq:daR}), $a=1/2T^{2}_{1}$ and $b=1/2T^{2}_{2}$ are parameters determined by pulse-duration of FH and SH pulses, while $\nu_{1}=1/v_{g3}-1/v_{g1}$ and $\nu_{2}=1/v_{g3}-1/v_{g2}$ are the so-called group-velocity mismatches. The pulse-duration term $\exp[-a(t+\xi\nu_{1})^{2}-b(t+\xi\nu_{2})^{2}]$ , as analyzed in Ref.\cite{Yishengyang}, is significantly crucial for ultrashort (e.g., picosecond, femtosecond, or even shorter) laser pulses. However, for phase-modulated broadband THG, pulse duration is generally around nanosecond \cite{Eimerl}, and the effect of that pulse-duration term is negligible.

 Since the pulse-duration term can be neglected, Eq.(\ref{eq:daR}) turns out to be the Fourier transformation of
 \begin{equation}\label{eq:phasemodulation}
    \exp\Big[i\sigma_{1}\sin\big(2\pi\Omega_{1}(t+\xi\nu_{1})\big)+i\sigma_{2}\sin\big(2\pi\Omega_{2}(t+\xi\nu_{2})\big)\Big].
 \end{equation}
 For simplicity, we initially apply Fourier transforms on $\exp\Big[i\sigma_{1}\sin\big(2\pi\Omega_{1}(t+\xi\nu_{1})\big)\Big]$, and achieve
 \begin{equation} \label{eq:fourier}
    \sum_{n=-\infty}^{\infty}J_{n}(\sigma_{1})\delta(\omega-2\pi {n}\Omega_{1})\exp(i\omega\xi\nu_{1}).
 \end{equation}
 Substituting (\ref{eq:fourier}) into (\ref{eq:A3amplitude}), results show that the intensity of TH pulse in frequency domain possesses the characteristic of
 \begin{equation} \label{eq:sincfunction}
    |{\tilde{A_{3}}(z,\omega)}|^{2} \propto sinc^{2} (\omega\nu_{1}z),
 \end{equation}
 which implies that the output TH pulse will become intensity-modulated if $\nu_{1} \neq {0}$. Since $\nu_{1} \neq {0}$ means no group-velocity mismatch between FH and TH pulses, we could conclude that the FM-to-AM conversion in process is basically caused by group-velocity mismatches between interacting phase-modulated pulses.

 Similarly, for Eq.(\ref{eq:phasemodulation}), we assume modulation frequencies of FH and SH pulses to be the same ($\Omega_{1}=\Omega_{2}=\Omega$), since random or unequal relations between $\Omega_{1}$ and $\Omega_{2}$ makes analysis complicated. Let $x=\sigma_{2}/\sigma_{1}$, and reorganize Eq.(\ref{eq:phasemodulation}) as below
 \begin{equation} \label{eq:expand}
     \exp i\sigma_{1}\left[\begin{array}{c}
     \sin(2\pi\Omega t)\big[\cos(2\pi\Omega\xi\nu_{1})+x \cos(2\pi\Omega\xi\nu_{2})\big]\\
     +\cos(2\pi\Omega t)\big[\sin(2\pi\Omega\xi\nu_{1})+x \sin(2\pi\Omega\xi\nu_{2})\big]
     \end{array}\right].
 \end{equation}
 Obviously, if group-velocity mismatches $\nu_{1}$ and $\nu_{2}$ satisfy
 \begin{equation} \label{eq:condition}
     \sin(2\pi\Omega\xi\nu_{1})+x\sin(2\pi\Omega\xi\nu_{2})=0,
 \end{equation}
 (\ref{eq:expand}) could be simplified to $\exp\big[i\sigma'\sin(2\pi\Omega t)\big]$, just with the new modulation depth changing to $\sigma'$ which is equivalent to $\sigma_{1}\big[\cos(2\pi\Omega\xi\nu_{1})+x \cos(2\pi\Omega\xi\nu_{2})\big]$. Compared with analysis ahead, this indicates that no amplitude modulation will be induced on output TH pulse.

 Under pump undepletion ($\xi\rightarrow 0$) and accordingly $\sin\theta\approx\theta$ approximations, by substituting $\nu_{1}$ and $\nu_{2}$, Eq.(\ref{eq:condition}) transforms to
 \begin{equation} \label{eq:relationship}
     \frac{1+x}{v_{g3}}=\frac{1}{v_{g1}}+\frac{x}{v_{g2}} \quad (x=\frac{\sigma_{2}}{\sigma_{1}}, \Omega_{2}=\Omega_{1}).
 \end{equation}
 This expression about group-velocity gives guidance for suppression of FM-to-AM conversion occurred in phase-modulated THG process. Coincidentally, Eq.(\ref{eq:relationship}) looks exactly the same with Eq.(19) in Ref.\cite{Yishengyang}, the only difference between them is the physical meaning of parameter $x$. $x$ here represents the ratio of modulation depth of SH to FH pulse, while $x$ in Ref.\cite{Yishengyang} represents the ratio of pulse duration of FH to SH pulse. Meanwhile, on the other hand, the two equations are consistent with each other, as frequency bandwidth of sinusoidally phase-modulated pulse is determined by $2\sigma\Omega$, while bandwidth of transform-limited ultrashort pulse is determined by the reciprocal of pulse duration.

 Based on the split-step Fourier transform and the fourth-order Runge-Kutta algorithm, the THG process in Type II potassium dihydrogen phosphate (KDP) crystal is numerically simulated to verify Eq.(\ref{eq:relationship}). In simulation, we change value of $v_{g3}$ to have different $x$ according Eq.(\ref{eq:relationship}), while values of $v_{g1}$ and $v_{g2}$ are calculated according the Sellmeier equations of KDP crystal \cite{Kirby}, as shown in Table I, where values of other basic input parameters of pulses are also presented.

 Fig. 1 plot temporal profiles of output three pulses in case of different $x$, with values of $\sigma_{1}$ and $\sigma_{2}$ are fixed to be $15$ and $30$, respectively. Fig. $2$ gives practical temporal profiles of output pulses in KDP crystal ($v_{g3}=1.92\times10^8 m/s$). Compared with the severe intensity modulations in Fig. $2$, intensity modulation of Fig. 1 are much smaller, especially when $x=\sigma_{2}/\sigma_{1}$ is satisfied (Fig. 1b). When $x$ is away from the ratio $\sigma_{2}/\sigma_{1}$, intensity modulations gets more severe. All these properties reveal that there will be no AM induced on output pulses when $x$ matches with the ratio of $\sigma_{2}/\sigma_{1}$, which confirms the validity of Eq.(\ref{eq:relationship}).

 However, see Fig. 1b closely, there are still some residual intensity modulations on temporal profiles. This is because the approximation of $\sin\theta\approx\theta$ is assumed for Eq.(\ref{eq:condition}). If we set $\sigma_{1}=\sigma_{2}$, then we can still achieve Eq.(\ref{eq:relationship}) exactly without the assumption of $\sin\theta\approx\theta$. Fig. $3$ presents temporal profiles of output three pulses with $\sigma_{1}=\sigma_{2}=15$ and $x=1$, which shows no intensity modulation.

 Now we could explain the phenomenon that the FM-to-AM conversion can be suppressed at the retracing point of a crystal from another aspect. As is known, the THG retracing point of a crystal corresponds to a special relationship of $\frac{3}{v_{g3}}=\frac{1}{v_{g1}}+\frac{2}{v_{g2}}$, which is identical with Eq.(\ref{eq:relationship}) when $\sigma_{2}=2\sigma_{1}$ is chosen \cite{Yisheng}. Since finding a crystal with appropriate retracing point is somewhat difficult, according Eq.(\ref{eq:relationship}), suppression of FM-to-AM can be achieved easily just by tuning the ratio of modulation depth of FH to SH pulse, which we think is more feasible in practical.\\[1ex]

 This work was partially supported by the National Natural Science Foundation of China (Grant No. 60708007), and the Science and Technology Foundation of Chinese State Key Laboratory of High Temperature and Density Plasma Physics (Grant No. 9140C6803010802).

 \begin{figure}
     \begin{center}
         \includegraphics[width=0.5\textwidth]{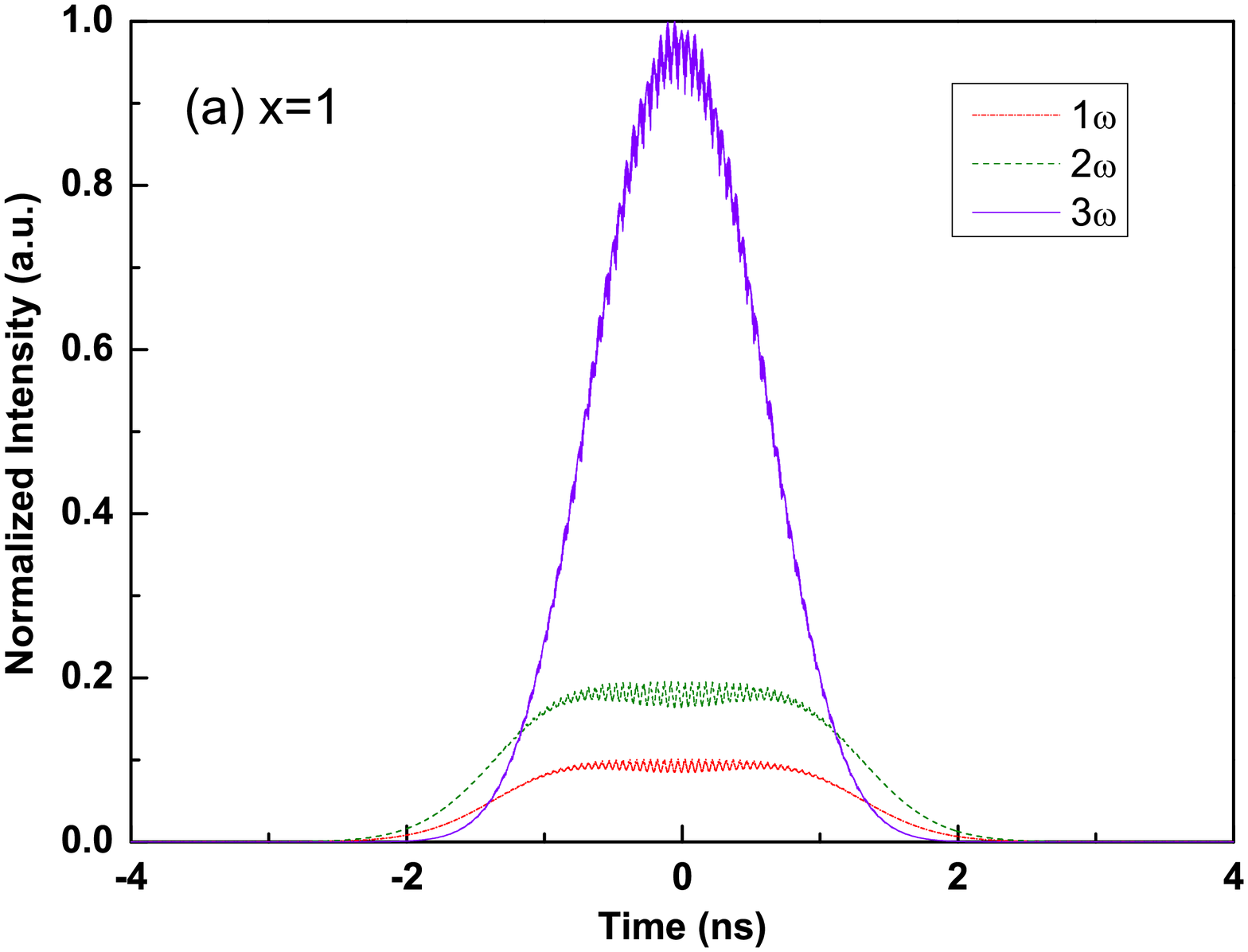}\\
         \includegraphics[width=0.5\textwidth]{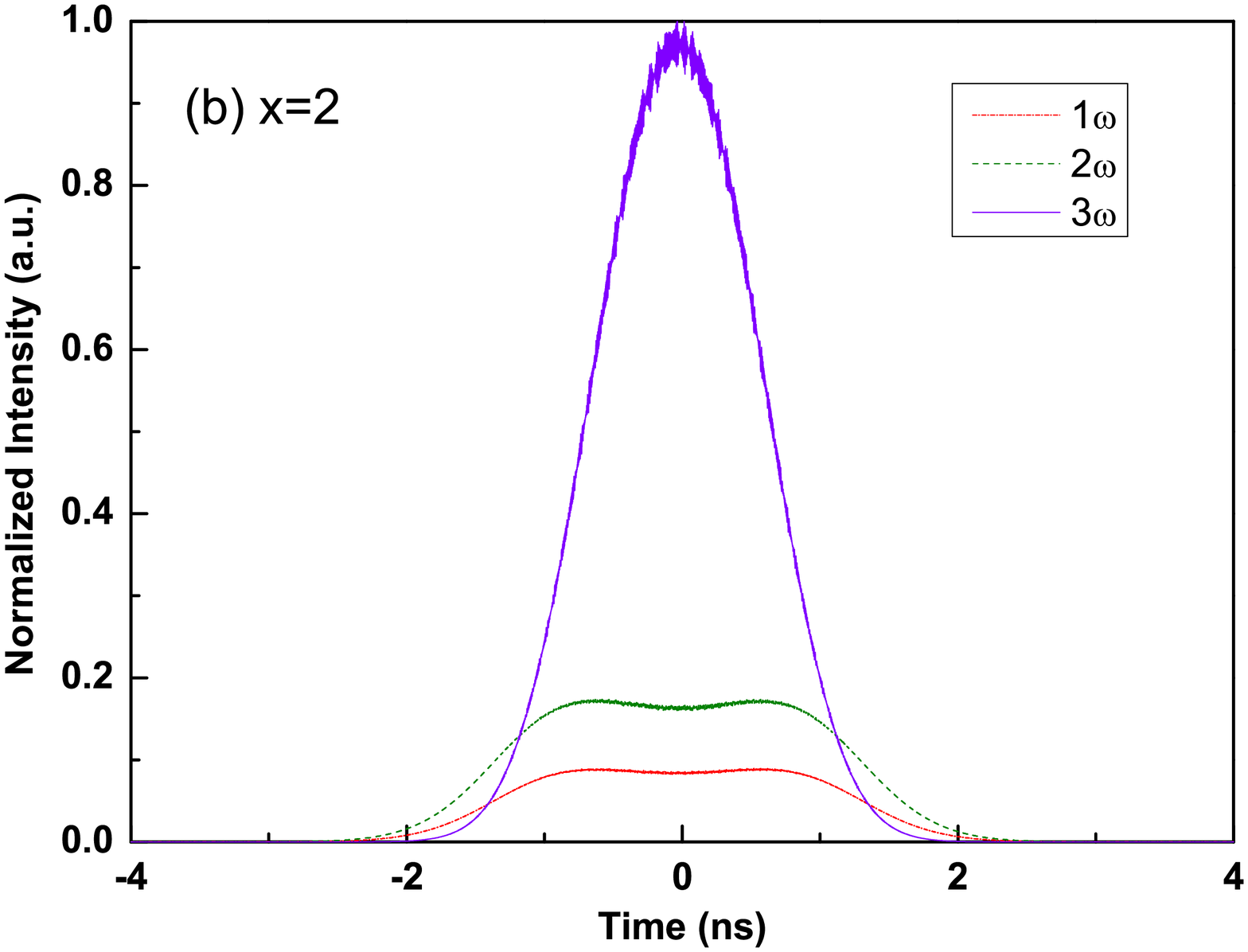}\\
         \includegraphics[width=0.5\textwidth]{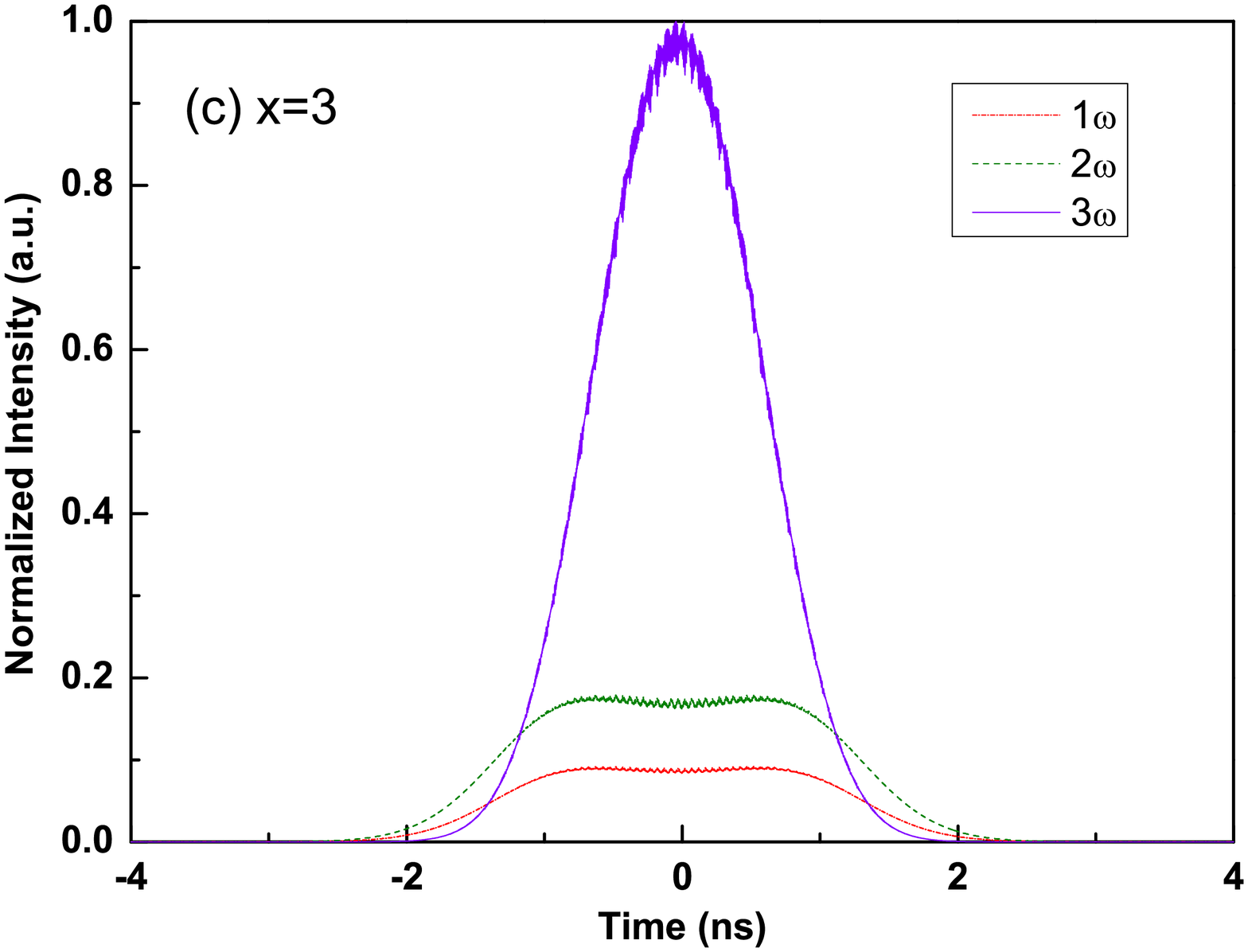}\\
         \includegraphics[width=0.5\textwidth]{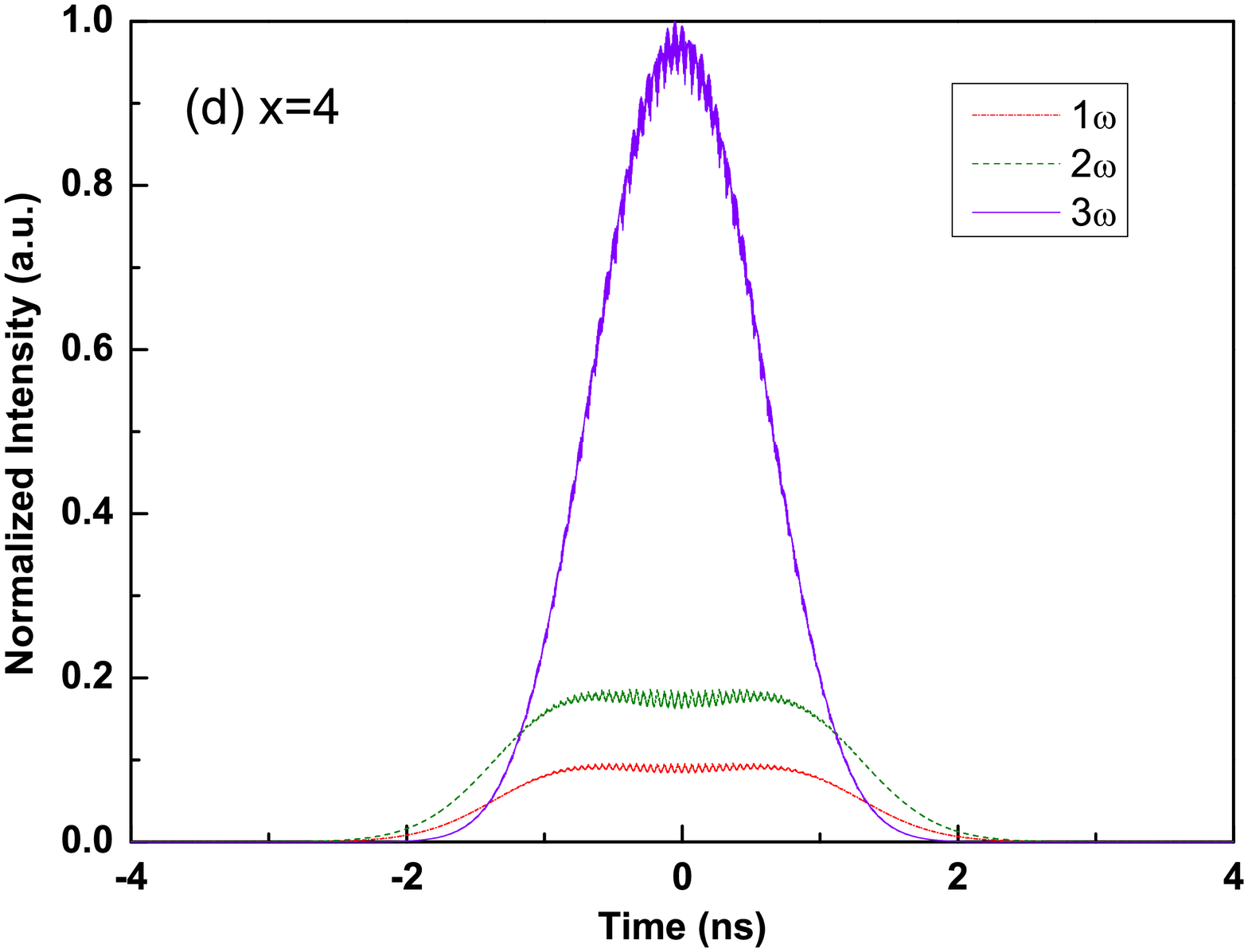}\\
         \caption{Temporal profiles with different $x$}
     \end{center}
 \end{figure}

 \begin{figure}
     \begin{center}
         \includegraphics[width=0.5\textwidth]{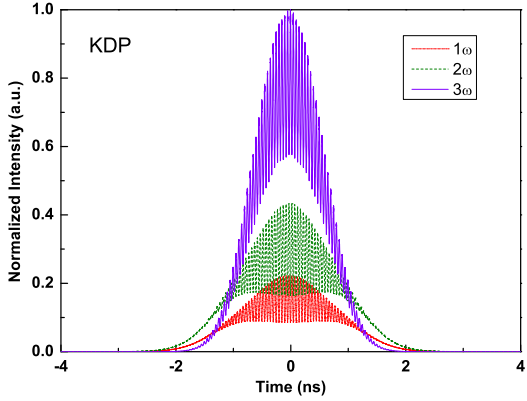}
         \caption{Practical temporal profiles in KDP crystal}
     \end{center}
 \end{figure}

 \begin{figure}
     \begin{center}
         \includegraphics[width=0.5\textwidth]{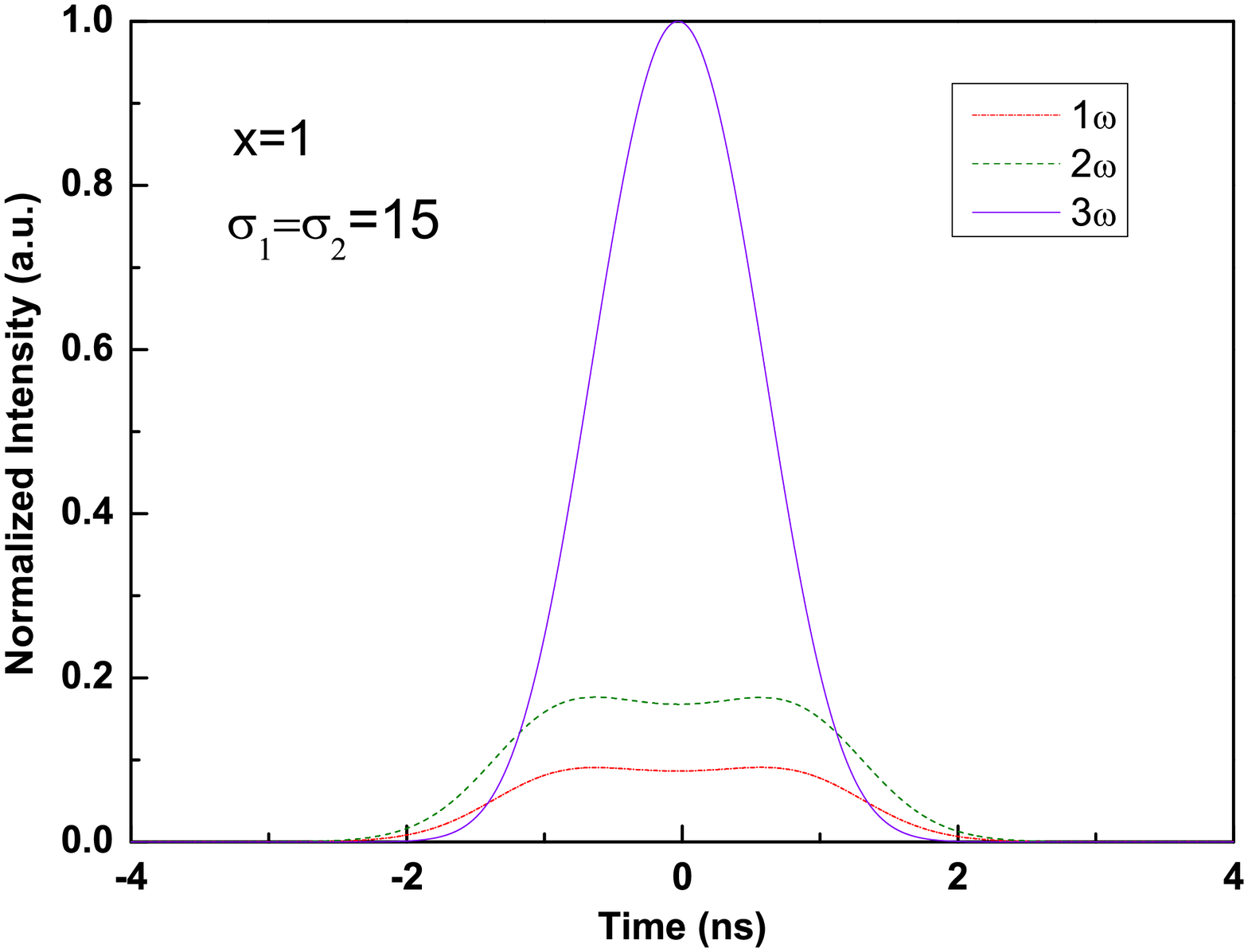}
         \caption{Temporal profiles with $x=1$ and $\sigma_1=\sigma_2=15$}
     \end{center}
 \end{figure}

\begin{table}
     \caption{Basic Parameters of Input $1\omega$ and $2\omega$ pulses}
     \begin{tabular}{lcccc} \hline
         Wavelength  & Pulse Duration & Modulation Frequency & Group-velocity            & Peak Intensity\\
         \           & $T_{0}[ns]$    & $\Omega[GHz]$        & in KDP [m/s]              & $I_{0}[GW/cm^2]$ \\[1ex] \hline
         $1\omega/1.053 \mu m$ & 1    & 10                   & 2.02$\times10^8$          & 1       \\
         $2\omega/0.527 \mu m$ & 1    & 10                   & 1.94$\times10^8$          & 2       \\ \hline %
     \end{tabular}
\end{table}


\begin{thebibliography}{20}
   \bibitem[1]{Wegner} P. J. Wegner, M. A. Henesian, D. R. Speck, C. Bibeau, R. B. Ehrlich, C. W. Laumann, J. K. Lawson, and T. L. Weiland,  Appl. Opt. \textbf{31}, 6414 (1992)
   \bibitem[2]{Photolith} M. Chen, Y. Chen, W. Hsiao, and Z. Gu, Thin Solid Films \textbf{515}, 8515 (2007).
   \bibitem[3]{Eimerl} D. Eimerl, J. M. Auerbach, C. E. Barker, D. Milam, and P. W. Milonni, Opt. Lett. \textbf{22}, 1208 (1997).
   \bibitem[4]{Hocquet} S. Hocquet, D. Penninckx, E. Bordenave, C. Gouedard, and Y. Jaouen, Appl. Opt. \textbf{47}, 3338 (2008).
   \bibitem[5]{Marozas} J. A. Marozas, J. Opt. Soc. Am. B \textbf{19}, 75 (2002).
   \bibitem[6]{Skupsky} S. Skupsky, R. W. Short, T. Kessler, R. S. Craxton, S. Letzring, and J. M. Soures, J. Appl. Phys. \textbf{66}, 3434926 (1989).
   \bibitem[7]{Hocquetsecond} S. Hocquet, G. Lacroix, and D. Penninckx, Appl. Opt. \textbf{48}, 2515 (2009).
   \bibitem[8]{Vidal} S. Vidal, J. Luce, and D. Penninckx, Opt. Lett. \textbf{36}, 3494 (2011).
   \bibitem[9]{Vidalsecond} S. Vidal, J. Luce, and D. Penninckx, Opt. Lett. \textbf{36}, 88 (2011).
   \bibitem[10]{Huabao} H. Cao, X. Lu, L. Li, X. Yin, W. Ma, J. Zhu, and D. Fan, Appl. Opt. \textbf{50}, 3609 (2011)
   \bibitem[11]{CHEN} Chen Y., Qian L., Zhu H., Fan D., Chin. Phys. Lett. \textbf{28}, 044209 (2011)
   \bibitem[12]{Wangwei} W. Wang, W. Han, F. Wang, J. Wang, L. Zhou, H. Jia, Y. Xiang, K. Li, F. Li, L. Wang, W. Zhong, X. Zhang, S. Zhao, and B. Feng, J. Opt. Soc. Am. B \textbf{28}, 475 (2011)
   \bibitem[13]{Yisheng} Y. Yang, B. Feng, W. Han, W. Zheng, F. Li, and J. Tan, Opt. Lett. \textbf{34}, 3848 (2009)
   \bibitem[14]{Armstrong} J. A. Armstrong, N. Bloembergen, J. Ducuing, and P. S. Pershan, Phys. Rev. \textbf{127}, 1918 (1962).
   \bibitem[15]{Yishengyang} Y. Yang, W. Han, W. Zheng, J. Tan, F. Li, F. Wang, Y. Xiang, K. Li, B. Feng, H. Jia, D. Cao, and J. Dong, Phys. Rev. A \textbf{78}, 053801 (2008).
   \bibitem[16]{Kirby} K. W. Kirby and L. G. DeShazer, J. Opt. Soc. Am. B \textbf{4}, 1072 (1987).

\end{thebibliography}
\end{document}